\documentclass[twocolumn,secnumarabic,amssymb, amsmath, nofootinbib,tightenlines,
nobibnotes, aps, prl,epsfig]{revtex4}
\usepackage{graphicx}
\usepackage{dcolumn}
\usepackage{bm}
\begin{document}
\preprint{APS/123-QED}
\title{The behavior of the heavy-quarks structure functions at small-$x$ }

\author{G.R.Boroun}%
 \email{grboroun@gmail.com; boroun@razi.ac.ir }
\author{B.Rezaei }
\altaffiliation{brezaei@razi.ac.ir}
\affiliation{ Physics Department, Razi University, Kermanshah
67149, Iran}
\date{\today}
\begin{abstract}
The behavior of the charm and bottom structure functions
($F_{k}^{i}(x,Q^{2})$, i=c,b; k=2,L) at small-$x$ is considered
with respect to the hard-Pomeron and saturation models. Having
checked that this behavior predicate the heavy flavor reduced
cross sections concerning the unshadowed and shadowed corrections.
We will show that the effective exponents for the unshadowed and
saturation corrections are independent of $x$ and $Q^{2}$, and
also the effective coefficients are dependent to $\ln{Q^{2}}$
compared to Donnachie-Landshoff (DL) and color
dipole (CD) models.\\
\end{abstract}
\maketitle
\centerline{\textbf{1. Introduction}}

In perturbative quantum chromodynamics (PQCD) calculations the
production of heavy quarks at HERA proceeds dominantly via the
direct boson- gluon fusion (BGF)(Fig.1), where the photon
interacts indirectly with a gluon in the proton by the exchange of
a heavy quark pair [1-10]. In the BGF dynamic, the charm(bottom)
quark is treated as a heavy quark and its contribution is realized
by fixed- order perturbative theory. As to the measurements of
HERA [11-15], the charm (bottom) contribution to the structure
function at small $x$ is a large fraction of the total, since this
value is approximately $30\%$ ($1\%$) fraction of the total. This
behavior  directly is related to the growth of the gluon
distribution at small $x$. We know that the gluons couple only
through the strong interaction, consequently the gluons are not
directly probed in DIS. The only way to indirect contribution is
via the
 $g{\rightarrow}q_{H}\bar{q}_{H}~(q_{H}=c,b)$
transition. This involves the computation of the BGF process
$\gamma^{\star}g{\rightarrow}q_{H}\bar{q}_{H}$. This process can
be created when the squared invariant mass of the hadronic final
state has the condition that runs as follows $W^{2}{\geq}4m_{q_{H}}^{2}$.\\
In the framework of DGLAP (Dokshitzer-
Gribov-Lipatov-Altarelli-Parisi) [16-18] dynamics, considering the
heavy-flavour physics is in the framework of
variable-flavor-number scheme (VFNS) [19,20]. In this scheme, the
mass logarithms are resummed through all orders into a heavy quark
density according to the DGLAP evolution equations. All
logarithmic terms of the heavy flavor Wilson coefficients are
obtained by factorization of this quantity into the massive
operator matrix elements. The study of heavy flavor production can
be done in deep inelastic electron-proton scattering, which was
investigated experimentally at HERA and recently at LHC. These
data for heavy quark production, have been proposed in the
framework of the fixed-flavor-number scheme (FFNS)(where only
light degrees of freedom are considered as active), that it was
calculated for $F_{2}^{c}$ and $F_{2}^{b}$
[11-15].\\
The heavy flavor structure functions $F_{k}^{i}(x,Q^{2})$($k=2,L$
and $i=c,b$ ) are dependent to the parton distribution functions.
In the small-$x$ range, further simplification is obtained by
neglecting the contributions caused by incoming light quarks and
antiquarks. This is justified because they vanish at LO and are
numerically suppressed at NLO for small values of $x$, while the
gluon contribution is a matter at this region [6-7]. In axial
gauges, the leading double logarithms (i.e. $\ln(Q^{2})\ln(1/x)$)
are generated by ladder diagrams in which the emitted gluons have
strongly been ordered with respect to the transverse and
longitudinal momenta. The sum of theses momenta predicts that the
gluon distribution increases as $x$ decreases. Clearly this
increase cannot go on indefinitely. When the density of gluons
becomes too large, they can no longer be treated as free partons
[21-22]. At very small $x$ we expect annihilation or recombination
of gluons to occur and these shadowing corrections give rise to
nonlinear terms in the evolution equation for the gluon
distribution function. This picture allows us to write the GLR-MQ
(Gribov, Levin, Ryskin, Mueller and Qiu) equation for the gluon
distribution function
 at small-$x$ [23-24]. We expect the gluon correlations
at small-$x$ to tame the behavior of the gluon distribution
function. Therefore we observe that the heavy flavor structure
functions (HFSFs) and also heavy reduced cross
sections behaviors are tamed  the saturation effects.\\
An important point in gluon saturation approach is the
$x$-dependent saturation scale $Q_{s}^{2}(x)$, where it is the
critical line between non-linear and linear effects and it is an
intrinsic characteristic of a dense gluon system
[25-26].\\
In this paper, we also investigate this non-linear behavior for
the charm and bottom quarks at small-$x$ related to the GLR-MQ
evolution equation. Then we will apply the geometric scaling
parameterization
 in according with the critical line $Q^{2}=Q_{s}^{2}(x)$. We will do this, because the geometric scaling of the dipole cross section
 gives the similar scaling of the
 quantity $\alpha_{s}(Q^{2})\frac{xg(x,Q^{2})}{Q^{2}}$ that is dominant in the charm and bottom structure functions.\\
 The content of our paper is as follows. In the next section we
 give  a summary about heavy quark structure functions and color dipole model with starting gluon distribution along the critical line.
  Then we will study  the heavy quark structure functions for $Q^{2}{\geq}Q_{s}^{2}(x)$ in sections 3-5,
  respectively. In Sec.6 we present the behavior of the HFSFs exponents
  at unshadowed and shadowed corrections to the gluon density
  behavior and also in geometrical scaling. Finally we give our
  conclusions in Sec.7.\\

\centerline{\textbf{2. A short theoretical input}}

 In this section we briefly present the theoretical part of our
 analysis. The reader can be referred to the Refs.[27-36] for more details.\\
The heavy quark contributions $F_{k}^{i}(x,Q^{2},m^{2}_{i})$ to
the proton structure function at small-$x$ (where only the gluon
contributions are considerable) are given by this form
\begin{eqnarray}
F_{k}^{i}(x,Q^{2},m^{2}_{i})&=&2e_{i}^{2}\frac{\alpha_{s}(\mu^{2})}{2\pi}\int_{1-\frac{1}{a}}^{1-x}dzC_{g,k}^{i}
(1-z,\zeta)\nonumber\\
&& {\times}G(\frac{x}{1-z},\mu^{2}),
\end{eqnarray}
where $a=1+4\zeta(\zeta{\equiv}\frac{m_{i}^{2}}{Q^{2}})$, $G=xg$
is the gluon distribution function and $\mu^{2}$ denotes the
factorization scale. Here $C^{i}_{g,k}$ are the heavy coefficient
functions in BGF at LO and NLO analysis and in the NLO analysis
\begin{eqnarray}
\alpha_{s}(\mu^{2})=\frac{4{\pi}}{\beta_{0}ln(\mu^{2}/\Lambda^{2})}
-\frac{4\pi\beta_{1}}{\beta_{0}^{3}}\frac{lnln(\mu^{2}/\Lambda^{2})}{ln(\mu^{2}/\Lambda^{2})}
\end{eqnarray}
with $\beta_{0}=11-\frac{2}{3}n_{f},
\beta_{1}=102-\frac{38}{3}n_{f} $ ($n_{f}$ is the number of active
flavors).\\
At small-$x$, perturbative QCD predicts an increase in the gluon
distribution  tamed by saturation effects. The physical picture of
this process is provided in the rest frame of the proton. In the
small $x$ limit, the virtual photon splits into a $q\overline{q}$
color dipole followed by the interaction of this dipole with the
color fields in the proton. The dipole cross section has been
defined by [37]
\begin{eqnarray}
\sigma_{dipole}(x,\mathbf{r})=\sigma_{0}(1-e^{-\mathbf{r}^2Q_{s}^{2}(x)/4}),
\end{eqnarray}
in which $r$ is the transverse separation of the $q\overline{q}$
pair and $Q_{s}(x)$ parameterized as
$Q_{s}^{2}(x)=Q_{0}^{2}(x_{0}/x)^{\lambda}~ GeV^{2}$. The
important property of the dipole cross section is its geometric
scaling (GS), which is a well-known property of DIS for small-$x$
values [38-41]. Therefore the proton cross section is dependent
upon the single variable $\tau=Q^{2}/Q_{s}^{2}(x)$, as
\begin{eqnarray}
\sigma_{\gamma^{*} p}(x,Q^{2})=\sigma_{\gamma^{*} p}(\tau).
\end{eqnarray}\\

{\textbf{3. Linear behavior for the evolution of the HFSFs}}

The heavy flavor structure functions (HFSFs) are described as
Mellin convolutions between the gluon distribution $G(x,\mu^{2})$
and the Wilson coefficients $C_{k,g}^{i}(x,\frac{Q^{2}}{\mu^{2}})$
as
\begin{eqnarray}
F_{k}^{i}(x,Q^{2})=C_{k,g}^{i}(x,\frac{Q^{2}}{\mu^{2}}){\otimes}G(x,\mu^{2}),~~(k=2
\& L, i=c \& b)
\end{eqnarray}
here the Mellin convolutions is given by
\begin{eqnarray}
f_{1}(x){\otimes}f_{2}(x){\equiv}\int_{x}^{1}\frac{dy}{y}f_{1}(y)f_{2}(\frac{x}{y}).
\end{eqnarray}
The $Q^{2}$ evolution equation for the HFSFs (Eq.5) is expressed
in terms of these structure functions, as we have it:
\begin{eqnarray}
\frac{{\partial}F^{i}_{k}(x,Q^{2})}{{\partial}\ln
Q^{2}}&=&\sum_{n=1}n\frac{{d}{\ln}\alpha_{s}}{{d}\ln
Q^{2}}F^{i,(n)}_{k}(x,Q^{2})+P_{gg}{\otimes}F^{i}_{k}(x,Q^{2})\nonumber\\
&&+\frac{d{\ln}C_{k,g}^{i}}{d{\ln
}Q^{2}}{\otimes}F^{i}_{k}(x,Q^{2}),
\end{eqnarray}
in which the corresponding physical evolution kernel $P_{gg}$ can
be derived from DGLAP evolution equation at small-$x$ as it
follows:
\begin{equation}
\frac{{\partial}G(x,\mu^{2})}{{\partial}{\ln}Q^{2}}=P_{gg}(x,\alpha_{s}){\otimes}
G(x,\mu^{2}).
\end{equation}
This kernel is corresponding to the massless Wilson coefficients
in leading order (LO) up to next-to-next-to leading order (NNLO)
[42-44], and also the heavy contributions are in leading order and
next-to-leading order by using massive Wilson coefficients in the
asymptotic region $Q^{2} >> m^{2}$.\\
According to the Regge pole approach, the distribution functions
can be controlled by Pomeron exchange at small $x$, since these
behaviors are correspondent to the BFKL
(Balitskii-Fadin-Kuraev-Lipatov) Pomeron[45-48] ideas as extended
by adding a hard Pomeron  which describes the small-$x$ HERA data
up to $Q^{2}$  of a few hundred GeV$^{2}$ values. The small-x
asymptotic behavior for the gluon  and heavy  flavors can be
exploited to the evolution equations of the HFSFs. Therefore,
linear evolution of the HFSFs at small-$x$ can be found as
\begin{eqnarray}
F_{k}^{i}(x,Q^{2})|_{x{\rightarrow}0}=\sum_{n=1}F^{i,(n)}_{k}(x,Q_{0}^{2})I_{k}^{i,(n)},
\end{eqnarray}
where
\begin{eqnarray}
I_{k}^{i,(n)}&=&\exp(\int_{Q_{0}^{2}}^{Q^{2}}{d}\ln
Q^{2}(\sum_{n=1}n\frac{{d}{\ln}\alpha_{s}}{{d}\ln
Q^{2}}+P_{gg}{\otimes}x^{\lambda}\nonumber\\
&&+\frac{{d}{\ln}C_{k,g}^{i}}{{d}\ln Q^{2}}{\otimes}x^{\lambda})).
\end{eqnarray}
Now, we have a compact small-$x$ formula for the heavy flavor
ratio $R^{i}=\frac{F_{L}^{i}}{F_{2}^{i}}$ which greatly simplifies
the extraction of $F^{i}_{2}$ from measurements of reduced heavy
cross sections $\sigma_{r}^{i}$:
\begin{eqnarray}
\sigma_{r}^{i}(x,Q^{2})=
F^{i}_{2}(x,Q^{2})[1-\frac{y^{2}}{1+(1-y)^{2}}R^{i}(x,Q^{2})].
\end{eqnarray}
Therefore, we found a small-$x$ formula for the ratio $R^{i}$ by
the following form
\begin{eqnarray}
R^{i}(x,Q^{2})&=&\frac{F_{L}^{i}(x,Q^{2})}{F_{2}^{i}(x,Q^{2})}=\frac{F^{i}_{L}(x,Q_{0}^{2})}{F^{i}_{2}(x,Q_{0}^{2})}{\times}\\
&&\exp(\int_{Q_{0}^{2}}^{Q^{2}}{d}\ln
Q^{2}(\frac{{d}[{\ln}C_{L,g}^{i}-{\ln}C_{2,g}^{i}]}{{d}\ln
Q^{2}}{\otimes}x^{\lambda})).\nonumber
\end{eqnarray}\\

{\textbf{4. Nonlinear behavior for the evolution of the HFSFs}}

As mentioned above the HFSFs linear evaluation equation is based
on a hard-Pomeron behavior for the gluon and heavy structure
functions [49-56]. The gluon density increases with decreasing $x$
and we must reach the region in which gluon-gluon interactions
confine the growth implied by this behavior concerning the gluon
distribution, $G \sim x^{-\lambda}$, as this behavior will violate
unitarity when $x{\rightarrow}0$. Thus, we  discuss absorption
effects which tame the violation of unitarity. At sufficient
small-$x$ values two gluons in different cascades may interact and
  the gluon ladders fusion are generally important. Therefore  the
gluon density is decreasing and shadowing contributions can no
longer be neglected [21-22,57-60]. Shadowing corrections, which
take into account the fusion of $t$-channel gluons, modify the
linear DGLAP equation for the gluon distribution by adding a
negative term proportional to quadratic in $g(x,Q^{2})$. This
picture allows us to write the GLR-MQ equation for the gluon
distribution function behavior at small-$x$ symbolically as:
\begin{eqnarray}
\frac{{\partial}xg(x,Q^{2})}{{\partial}{\ln}Q^{2}}&=&\frac{{\partial}xg(x,Q^{2})}{{\partial}{\ln}Q^{2}}|_{DGLAP}\nonumber\\
&&-\frac{81\alpha^{2}_{s}\gamma}{16R^{2}Q^{2}}\int\frac{dy}{y}[yg(y,Q^{2})]^{2}.
\end{eqnarray}
The nonlinear shadowing term, ${\propto}-[g]^{2}$, arises from
perturbative QCD diagrams which couple four gluons into two gluons
so that two gluon ladders recombine into a single one. The minus
sign occurs because the scattering amplitude corresponding to a
gluon ladder is predominantly imaginary .
 Thus the equation (13) becomes nonlinear in $xg$.
 We neglect the quark-gluon emission diagrams
due to their little importance on the rich gluon in small-$x$
region and we work under an approximation of neglecting
contribution
 from the high twist gluon distribution $G_{HT}(x,Q^{2})$ [24].\\
In what follows it is convenient to use directly the reduced gluon
distribution function behavior according to the Eq.13 as the
$Q^{2}$ evolution of the HFSFs modified by
\begin{eqnarray}
\frac{{\partial}F^{i}_{k}(x,Q^{2})}{{\partial}\ln
Q^{2}}&=&\frac{{\partial}F^{i}_{k}(x,Q^{2})}{{\partial}\ln
Q^{2}}[Eq.7]\nonumber\\
&&-2e_{i}^{2}\frac{\alpha_{s}}{2\pi}\frac{\alpha^{2}_{s}\gamma}{R^{2}Q^{2}}\int_{1-\frac{1}{a}}^{1-x}dzC_{g,k}^{i}
(1-z,\zeta)\nonumber\\
&&
{\times}\frac{G^{2}(\frac{x}{1-z},\mu^{2})}{2\lambda}[1-(\frac{\chi}{1-z})^{2\lambda}].
\end{eqnarray}
Here $\chi=\frac{x}{x_{0}}$, where $x_{0}(=0.01)$ is the boundary
condition that the gluon distribution joints smoothly the
unshadowed region and  $R$  is the radius of the proton. The first
term is the linear evolution (Eq.7) and the second term is due to
the 2-gluon density. Therefore the shadowing corrections to the
HFSFs can be defined by Eq.14. To obtain a differential equation
for shadowing corrections to the HFSFs, we write out the sum and
the coefficient  function explicitly. Consequently we find an
inhomogeneous first-order differential equation which determines
 HFSFs shadowing corrections  in terms of shadowed gluons. Eq.
(14) can be rewritten in the following form:
\begin{eqnarray}
\frac{{\partial}F^{i}_{k}(x,Q^{2})}{{\partial}\ln
Q^{2}}-\eta(x,Q^{2}) F^{i}_{k}(x,Q^{2})=-S^{i}_{k}(x,Q^{2}),
\end{eqnarray}
where
\begin{eqnarray}
S^{i}_{k}(x,Q^{2})=2e_{i}^{2}\frac{\alpha_{s}}{2\pi}\frac{\alpha^{2}_{s}\gamma}{R^{2}Q^{2}}\int_{1-\frac{1}{a}}^{1-x}dzC_{g,k}^{i}
(1-z,\zeta)\nonumber\\
{\times}\frac{G^{2}(\frac{x}{1-z},\mu^{2})}{2\lambda}[1-(\frac{\chi}{1-z})^{2\lambda}],
\end{eqnarray}
and
\begin{eqnarray}
\eta(x,Q^{2})=(\sum_{n=1}n\frac{{d}{\ln}\alpha_{s}}{{d}\ln
Q^{2}}+P_{gg}{\otimes}x^{\lambda}+\frac{d{\ln}C_{k,g}^{i}}{d{\ln
}Q^{2}}{\otimes}x^{\lambda}).\nonumber\\
\end{eqnarray}
 The general solution of
Eq. (15) where tames the behavior of the HFSFs at small-$x$ has
the following form:
\begin{eqnarray}
F^{i}_{k}(x,Q^{2})&=&e^{\int_{Q_{0}^{2}}^{Q^{2}} \eta(x,Q^{2})d\ln
Q^{2}}{\times}[F^{i}_{k}(x,Q_{0}^{2})\nonumber\\
&&-\int_{Q_{0}^{2}}^{Q^{2}}S^{i}_{k}(x,Q^{2})e^{\int
-\eta(x,Q^{2})d\ln Q^{2}}d \ln Q^{2}].\nonumber\\
\end{eqnarray}
Therefore, shadowing corrections to the HFSFs modify the heavy
reduced cross section and also the ratio of the HFSFs, as we will
have:
\begin{eqnarray}
\sigma_{r}^{i}(x,Q^{2})&=&
F^{i}_{2}(x,Q^{2})[Eq.18][1-\frac{y^{2}}{1+(1-y)^{2}}\nonumber\\
&&{\times}R^{i}(x,Q^{2})[Eq.20]],
\end{eqnarray}
where
\begin{widetext}
\begin{eqnarray}
R^{i}(x,Q^{2})&=&\frac{F^{i}_{L}(x,Q^{2})}{F^{i}_{2}(x,Q^{2})}=[\frac{F^{i}_{L}(x,Q_{0}^{2})
-\int_{Q_{0}^{2}}^{Q^{2}}S^{i}_{L}(x,Q^{2})e^{\int
-\eta_{L}(x,Q^{2})d\ln Q^{2}}d \ln Q^{2}}{F^{i}_{2}(x,Q_{0}^{2})
-\int_{Q_{0}^{2}}^{Q^{2}}S^{i}_{2}(x,Q^{2})e^{\int
-\eta_{2}(x,Q^{2})d\ln Q^{2}}d \ln Q^{2}}]\nonumber\\
&&{\times} \exp(\int_{Q_{0}^{2}}^{Q^{2}}{d}\ln
Q^{2}(\frac{{d}[{\ln}C_{L,g}^{i}-{\ln}C_{2,g}^{i}]}{{d}\ln
Q^{2}}{\otimes}x^{\lambda})).
\end{eqnarray}
\end{widetext}
Equations 18-20 satisfy the requirements expected for shadowed
distributions of the heavy quarks. These equations reduced to the
unshadowed distributions (Eqs.9-12) when shadowing is negligible;
that is, when $S^{i}_{k}(x,Q^{2}){\rightarrow}0$ which implies
$G_{sat}{\rightarrow}\infty$. Finally the DGLAP+GLRMQ evolution
joins smoothly into the DGLAP evolution when
$x{\rightarrow}x_{0}$.\\

{\textbf{5. Geometrical scaling of the  HFSFs}}

 In the saturation model [61], the dipole cross section is bounded by
 an energy independent value $\sigma_{0}$ (Eq.3) which imposes the
 unitarity condition, $\sigma_{q\overline{q}}{\leq}\sigma_{0}$
 with respect to the free parameters in the model[38-41]. These free
 parameters can be extracted from data within some specific models
 of DIS. In the Golec-Biernat-W$\ddot{u}$sthoff (GBW) model we see,
 $\sigma_{0}=23~ mb$, $\lambda\cong 0.3$, $Q_{0}=1~GeV/c$ and $x_{0}=3
 {\times}10^{-4}$. The dipole cross section for a small
 $q\overline{q}$ dipole is related to the gluon density at scale
 $\mu^{2}$ as:
 \begin{eqnarray}
\sigma_{q\overline{q}}=\frac{\pi^{2}}{3}r^{2} \alpha_{s}
xg(x,\mu^{2}).
 \end{eqnarray}
For small $r{\ll}2R_{0}(x)$ in the saturation model (where
$R_{0}(x)$ is the saturation scale at small-$x$), the gluon
density found [38-41] by the following form:
 \begin{eqnarray}
xg(x,\mu^{2})=\frac{3}{4\pi^{2}\alpha_{s}}\frac{\sigma_{0}}{R_{0}^{2}(x)},
 \end{eqnarray}
where at the geometric scale  for the boundary
$Q^{2}=Q^{2}_{s}(x)$ we have:
\begin{eqnarray}
\frac{\alpha_{s}}{2\pi}xg(x,Q^{2}=Q^{2}_{s}(x))=r_{0}x^{-\lambda},
 \end{eqnarray}
with $r_{0}=\frac{3}{8{\pi^{3}}} {\sigma_{0}} {x_{0}^{\lambda}}$.
Also R.S.Thorne [62] used the relationship between the dipole
cross section and the unintegrated gluon distribution and showed
that the gluon distribution at fixed coupling has this behavior:
\begin{eqnarray}
xg(x,Q^{2})&=&\frac{3\sigma_{0}}{4{\pi^{2}}\alpha_{s}}(-Q^{2}e^{{-Q^{2}}(x/x_{0})^{\lambda}}+(x_{0}/x)^{\lambda}\nonumber\\
&&{\times}(1-e^{{-Q^{2}}(x/x_{0})^{\lambda}})).
\end{eqnarray}
In order to be able to study the formal heavy flavor production
limit, the Bjorken variable $x=x_{B}$ was modified into the one as
follows:
\begin{eqnarray}
x=x_{B}(1+\frac{4m^{2}_{hf}}{Q^{2}}),
\end{eqnarray}
when a pair of heavy quarks $H\overline{H}$ ($c\overline{c}$ or
$b\overline{b}$) are produced in the final state. Therefore the
gluon distribution can be evaluated for heavy quarks according to
Eq.24. This is consistent with a general-mass-variable flavor
number schemes (GM-VFNS) [63]. In this case, the parameters
obtained from the best fit were $\sigma_{0}=29.12~mb$,
$\lambda=0.277$ and $x_{0}=0.41{\times}10^{-4}$ [38-41]. Because
the data for the bottom component
$\sigma^{b}{\sim}\frac{F_{2}^{b}}{Q^{2}}$ are not suitable for
scaling analysis( since they contain two few points [11-15]),
therefore we used the charm parameters in our determination
for bottom structure functions.\\
Therefore the dipole picture at the geometric scaling is suitable
for HFSFs analysis . If the characteristic size of the
$q_{H}\overline{q}_{H}$-pair is much smaller than the saturation
radius, with decreasing $x$, it can  show that the behavior of the
HFSFs  at the scale $<\mu^{2}>$ is given by
\begin{eqnarray}
F_{k}^{i}(x,Q^{2})=r_{0}Q_{s}^{2}(x)[C_{k,g}^{i}(x,\frac{Q^{2}}{\mu^{2}}){\otimes}x^{\lambda}].
\end{eqnarray}

We summarized our results in sections 3-5 at Figs.2-4. In these
figures the results of calculation for charm and bottom structure
functions, longitudinal structure functions and reduced cross
sections are shown at $Q^{2}=12$ and $60~GeV^{2}$ respectively. We
determined the HFSFs and reduced cross sections in linear behavior
(unshadowed) when shadowing is neglected. It was also made clear
in nonlinear behavior (shadowed) at $R=5~GeV^{-1}$ where the
gluons are spread throughout the entire proton and at
$R=2~GeV^{-1}$ where the gluons are concentrated in hot-spots
within the proton. We observe that these behaviors for the heavy
quarks functions are tamed at small-$x$. The reduced cross
sections are determined at the average inelasticity  $<y>$ in
according with the H1 data. We compared our results with GJR
parameterization [64], the ZEUS and H1 data [11-15]. We can
observe that there are a well agreement between our unshadowed,
shadowed
and saturation results with the experimental data accompanied by total errors.\\

{\textbf{6. Heavy flavor exponents behavior}}

 In the double asymptotic limit, the behavior of the gluon
 distribution is expected to rise approximately as a power of $x$
 towards small-$x$. Because, the behavior of the HFSFs and heavy flavor reduced cross sections are
 dependent on the gluon  distribution behavior directly. Therefore we
 consider the power-like behavior of the HFSFs
 at small-$x$ as $f_{\xi}^{i}{\propto}x^{-\lambda_{\xi}^{i}(x,Q^{2})}$ where $f_{\xi}=F_{2}, F_{L}$ and $\sigma_{r}$. The logarithmic
 $x$-derivative of the heavy flavor functions concerning $\ln(\frac{1}{x})$
 can be defined [65] as:
 \begin{eqnarray}
H_{\xi}^{i}(x,Q^{2})=\frac{\partial
{\ln}f_{\xi}^{i}(x,Q^{2})}{\partial {\ln}(1/x)}.
 \end{eqnarray}
Here we would like to consider the relation between the effective
intercept $\lambda_{\xi}^{i}$ for heavy flavor functions and
logarithmic-$x$ derivatives of the heavy flavor functions where
 \begin{eqnarray}
H_{\xi}^{i}(x,Q^{2})= \lambda_{\xi}^{i}(x,Q^{2})+
{\ln}\frac{1}{x}\frac{\partial
\lambda_{\xi}^{i}(x,Q^{2})}{\partial {\ln}(1/x)}.
 \end{eqnarray}
For unshadowed heavy flavor functions (Figs.2-4) we used the
Pomeron intercept in our determinations, therefore the effective
intercept and $x$-slop strictly coincide with the hard Pomeron
intercept as we considered this behavior for the gluon
distribution at small-$x$. Therefore in this case, we will come
into the below:
 \begin{eqnarray}
H_{\xi}^{i}(Q^{2})= \lambda_{\xi}^{i}(Q^{2})=0.44.~~~(i=c,b)
 \end{eqnarray}
When we consider the saturation effects [25-26] on the charm
functions, the charm exponents obtained will be follows:
\begin{eqnarray}
 H_{\xi}^{c}(Q^{2})=
\lambda_{\xi}^{c}(Q^{2})=0.277.
 \end{eqnarray}
But this value is not coincident with the bottom functions, since
we do not have enough data for bottom component in scaling
analysis. One can see that exponents obtained for heavy flavor
functions at the shadowed region are dependent on the saturation
scale and $R$ parameter. For the nonlinear evaluation equation we
can observe that:
\begin{eqnarray}
H_{\xi}^{i}(x,Q^{2}){\neq} \lambda_{\xi}^{i}(x,Q^{2}).
 \end{eqnarray}
For heavy functions at $R=5$ and $2~GeV^{-1}$, there is not a
linear function when we fit it to all data because inequality (31)
is correct. For $R=5~GeV^{-1}$ there is a second-order function
and for $R=2~GeV^{-1}$ there is a three-order function. So we
conclude that at nonlinear evolution equations, the heavy
exponents are dependent on $x$ and $R$ values, for the shadowed
heavy flavors  are dependent on the values $x$, $Q^{2}$ and $R$,
i.e., ($\lambda_{\xi}^{i,
shadowed}=\lambda(x,Q^{2},R)$).\\
In addition to the $x$-slope, we want to consider the logarithmic
derivative of the heavy functions with respect to $\ln Q^{2}$, as
the $Q^{2}$-slopes are defined by:
 \begin{eqnarray}
\frac{\partial f_{\xi}^{i}(x,Q^{2})}{\partial {\ln}(Q^{2})}.
 \end{eqnarray}
Figure 5 shows the derivative as a function of $Q^{2}$ for
$x=0.001$ value. The derivative is observed to have a
logarithmically  approximate  rise with $Q^{2}$ for
$F_{2}^{i}(x,Q^{2})$. The ${\ln}(Q^{2})$ dependence of
$F_{2}^{i}(x,Q^{2})$ is observed to be non-linear. It can be well
described by a quadratic expression, since for each of heavy
quarks these derivatives can be described by the function
$b(x)+2c(x)\ln(Q^{2})$. The shape of these derivatives reflect the
behavior of the gluon distribution in the associated
kinematic range.\\
 This
scaling violation shows the transition from soft to hard dynamics.
We give this scaling violation to the function $\eta(Q^{2})$ then
we consider the heavy functions behavior as a fixed power of $x$:
\begin{eqnarray}
 f_{\xi}^{i}(x,Q^{2})=\eta_{\xi}^{i}(Q^{2})x^{-\lambda}.
 \end{eqnarray}
It would be implied that the Mellin transform
$f_{\xi}^{i}(j,Q^{2})$ would have a pole at $j=1+\lambda$, where
this singularity referred to as the hard-Pomeron singularity
[49-56]. We conclude that the power $\lambda^{,}$s in Eqs.29 and
30 are not dependent on $Q^{2}$ and  theses powers have got fixed
 values. In our determinations (Fig.5), each of theses
coefficient functions $\eta_{\xi}^{i}(Q^{2})$ vanished in a way
that $Q^{2}{\rightarrow}0$. In Fig.6 we compared our results for
$\eta_{2}^{c}(Q^{2})$ and $\eta_{2}^{b}(Q^{2})$ with DL model
[49-56] and color dipole model (CDM)
[1-4]. These results are comparable with others.\\

{\textbf{7. Conclusions}}

We have studied several aspects of the heavy flavors functions
behavior at small-$x$. To do this, we assumed that the heavy
parton density obeys power-laws having effective power. Therefore
we have applied the hard Pomeron and saturation methods in
obtaining the heavy flavors functions considering H1 and ZEUS
data. We have shown that geometrical scaling in DIS works well up
to the heavy flavors functions with a constant exponent. The merit
of this exponent is mainly due to its relation with the gluon
density at small-$x$. Results obtained  suggest that geometrical
scaling in heavy production is basically the same as for the
inclusive DIS. The value of shadowed exponents are different from
the one obtained for unshadowed corrections and geometrical
scaling in heavy functions. This difference is due to the
shadowing corrections on the heavy functions. Indeed the value of
the shadowed correction on the exponents depends on  how exactly
the gluon ladders couple to the proton or on how the gluons are
distributed within the proton, especially in hot-spot point. As a
result, the coefficient functions ($\eta(Q^{2})$) are dependent on
the $Q^{2}$ scale and the effective powers are constant. We
obtained our results for the $\eta(Q^{2})$
and compared them with DL and CD models.\\

{\textbf{Acknowledgment}}

The authors would like thank Y.Refahiyat
 for his careful revising of the paper with regard to its
 language.\\
 \newpage
\textbf{References}\\
1. N.N.Nikolaev and V.R.Zoller, Phys.Atom.Nucl\textbf{73},
672(2010).\\
2. N.N.Nikolaev and V.R.Zoller, Phys.Lett.B \textbf{509},
283(2001).\\
3. N.N.Nikolaev, J.Speth and V.R.Zoller, Phys.Lett.B\textbf{473},
157(2000).\\
4. R.Fiore,
N.N.Nikolaev and V.R.Zoller, JETP Lett\textbf{90}, 319(2009).\\
5.A.~V.~Kotikov, A.~V.~Lipatov, G.~Parente and N.~P.~Zotov Eur.\
Phys.\ J.\  C {\bf 26}, 51 (2002).\\
6. A.~Y.~Illarionov, B.~A.~Kniehl and A.~V.~Kotikov, Phys.\ Lett.\
B {\bf 663}, 66 (2008).\\
7. A. Y. Illarionov and A. V. Kotikov, Phys.Atom.Nucl. {\bf75}, 1234 (2012).\\
8. N.Ya.Ivanov, and B.A.Kniehl, Eur.Phys.J.C\textbf{59}, 647(2009).\\
9.  N.Ya.Ivanov, Nucl.Phys.B\textbf{814}, 142(2009).\\
10. J.Blumlein, et.al., Nucl.Phys.B\textbf{755}, 272(2006).\\
11. F.D. Aaron et al. [H1 Collaboration],Phys.Lett.b\textbf{665},
139(2008).\\
12. F.D. Aaron et al. [H1
Collaboration],Eur.Phys.J.C\textbf{65},89(2010).\\
13. H.Abramovicz et. al., [ZEUS Collaboration],
arXiv:hep-ex/1005.3396(2010).\\
14. H.Abramovicz et. al., [ZEUS Collaboration],
arXiv:hep-ex/1405.6915(2014).\\
15. H.Abramovicz et. al.,
[Combination H1 and ZEUS Collaboration], arXiv:hep-ex/1211.1182(2012).\\
16.Yu.L.Dokshitzer, Sov.Phys.JETP {\textbf{46}}, 641(1977).\\
17. G.Altarelli and G.Parisi, Nucl.Phys.B \textbf{126},
298(1977).\\
18. V.N.Gribov and L.N.Lipatov,
Sov.J.Nucl.Phys. \textbf{15}, 438(1972).\\
19. M.A.G.Aivazis, et.al., Phys.Rev.D\textbf{50},
3102(1994).\\
20. J.C.Collins, Phys.Rev.D\textbf{58},
094002(1998).\\
21. J.Kwiecinski, A.D.Martin and P.J.Sutton,
Phys.Rev.D\textbf{44}, 2640(1991).\\
22. J.Kwiecinski, A.D.Martin, R.G.Roberts and W.J.Stirling,
Phys.Rev.D\textbf{42},
3645(1990).\\
23. L.V.Gribov, E.M.Levin and M.G.Ryskin, Phys.Rep.\textbf{100},
 1(1983).\\
24. A.H.Mueller and J.Qiu, Nucl.Phys.B\textbf{268}, 427(1986).\\
25. K. Golec-Biernat, Acta.Phys.Polon.B\textbf{35}, No.12,
3103(2004).\\
26. F. Carvalho, et.al., Phys.Rev.C\textbf{79},
035211(2009).\\
27. M.Gluk, E.Reya and A.Vogt, Z.Phys.C\textbf{67}, 433(1995).\\
28. M.Gluk, E.Reya and A.Vogt, Eur.Phys.J.C\textbf{5}, 461(1998).\\
29. E.Laenen, S.Riemersma, J.Smith and W.L. van Neerven,
Nucl.Phys.B \textbf{392}, 162(1993).\\
30. S. Catani, M. Ciafaloni and F. Hautmann, Preprint
CERN-Th.6398/92, in Proceeding of the Workshop on Physics at HERA
(Hamburg, 1991), Vol. 2., p. 690.\\
31. S. Catani and F. Hautmann, Nucl. Phys. B \textbf{427},
475(1994).\\
32. S. Riemersma, J. Smith and W. L.
van Neerven, Phys. Lett. B \textbf{347}, 143(1995).\\
33. J.Kwiecinski and A.M.Stasto, Phys.Rev.D\textbf{66},
014013(2002).\\
34. A.M.Stasto, et.al., Phys.Rev.Lett\textbf{86},
596(2001).\\
35. E.Iancu, et.al., Phys. Lett. B\textbf{590}, 199(2004).\\
36. H.Kowalski and D.Teaney, Phys. Rev.D\textbf{68}, 114005(2003).\\
37. K. Golec-Biernat and M.Wusthoff, Phys.Rev.D\textbf{59},
014017(1998).\\
38. K. Golec-Biernat, J.Phys.G\textbf{28}, 1057(2002).\\
39. K. Golec-Biernat,Acta.Phys.Polon.B\textbf{33}, 2771(2002).\\
40. J.Bartles, et.al., Phys.Rev.D\textbf{66}, 014001(2002).\\
41. J.Bartles, et.al.,Acta.Phys.Polon.B\textbf{33},
2853(2002).\\
42. S.Moch and J.A.M.Vermaseren, Nucl.Phys.B\textbf{573},
853(2000).\\
43. S.Moch, J.A.M.Vermaseren and A.Vogt, Phys.Lett.B\textbf{606},
123(2005).\\
44.  J.A.M.Vermaseren, A.Vogt and S.Moch,,
Nucl.Phys.B\textbf{724},
3(2005).\\
45. E.A.Kuraev, L.N.Lipatov and V.S.Fadin, Phys.Lett.B
\textbf{60}, 50(1975).\\
46. E.A.Kuraev, L.N.Lipatov and V.S.Fadin, Sov.Phys.JETP
\textbf{44}, 433(1976).\\
47. E.A.Kuraev, L.N.Lipatov and V.S.Fadin, ibid. \textbf{45},
199(1977).\\
48. Ya.Ya.Balitskyii and L.N.Lipatov, Sov.J.Nucl.Phys. \textbf{28}, 822(1978).\\
49. A.Donnachie and P.V.Landshoff, Z.Phys.C \textbf{61},
139(1994).\\
50. A.Donnachie and P.V.Landshoff, Phys.Lett.B \textbf{518},
63(2001).\\
51. A.Donnachie and P.V.Landshoff, Phys.Lett.B \textbf{533},
277(2002).\\
52. A.Donnachie and P.V.Landshoff, Phys.Lett.B \textbf{470},
243(1999).\\
53. A.Donnachie and P.V.Landshoff, Phys.Lett.B \textbf{550}, 160(2002).\\
54. R.D.Ball and P.V.landshoff, J.Phys.G\textbf{26}, 672(2000).\\
55. J.R.Cudell, A.Donnachie and P.V.Landshoff, Phys.Lett.B \textbf{448}, 281(1999).\\
56. P.V.landshoff, arXiv:hep-ph/0203084 (2002).\\
57. K.J.Eskola, et.al., Nucl.Phys.B\textbf{660}, 211(2003).\\
58. K.Kutak and A.M.Stasto, Eur.Phys.J.C\textbf{41}, 343(2005).\\
59. M.Kazlov and E.Levin, Nucl.Phys. A\textbf{764}, 498 (2006).\\
60. M.A.Kimber, J.Kwiecinski and A.D.Martin,  Phys.Lett. B\textbf{508}, 58(2001).\\
61. K.Golec-Biernat, arXiv:hep-ph/0812.1523(2008).\\
62. R.S.Thorne, Phys.Rev.D\textbf{71}, 054024(2005).\\
63. G.Beuf, C.Royon and D.Salek, arXiv:hep-ph/0810.5082(2008).\\
64.M. Gluck, P. Jimenez-Delgado, E. Reya,
Eur.Phys.J.C\textbf{53},355(2008).\\
65. P.Desgrolard et.al., JHEP\textbf{02}, 029(2002).\\
\begin{figure}
\includegraphics[width=0.2\textwidth]{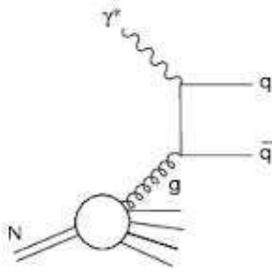}
\caption{ Boson-gluon-fusion (BGF) graph. }\label{Fig1}
\end{figure}
\begin{figure}
\includegraphics[width=1\textwidth]{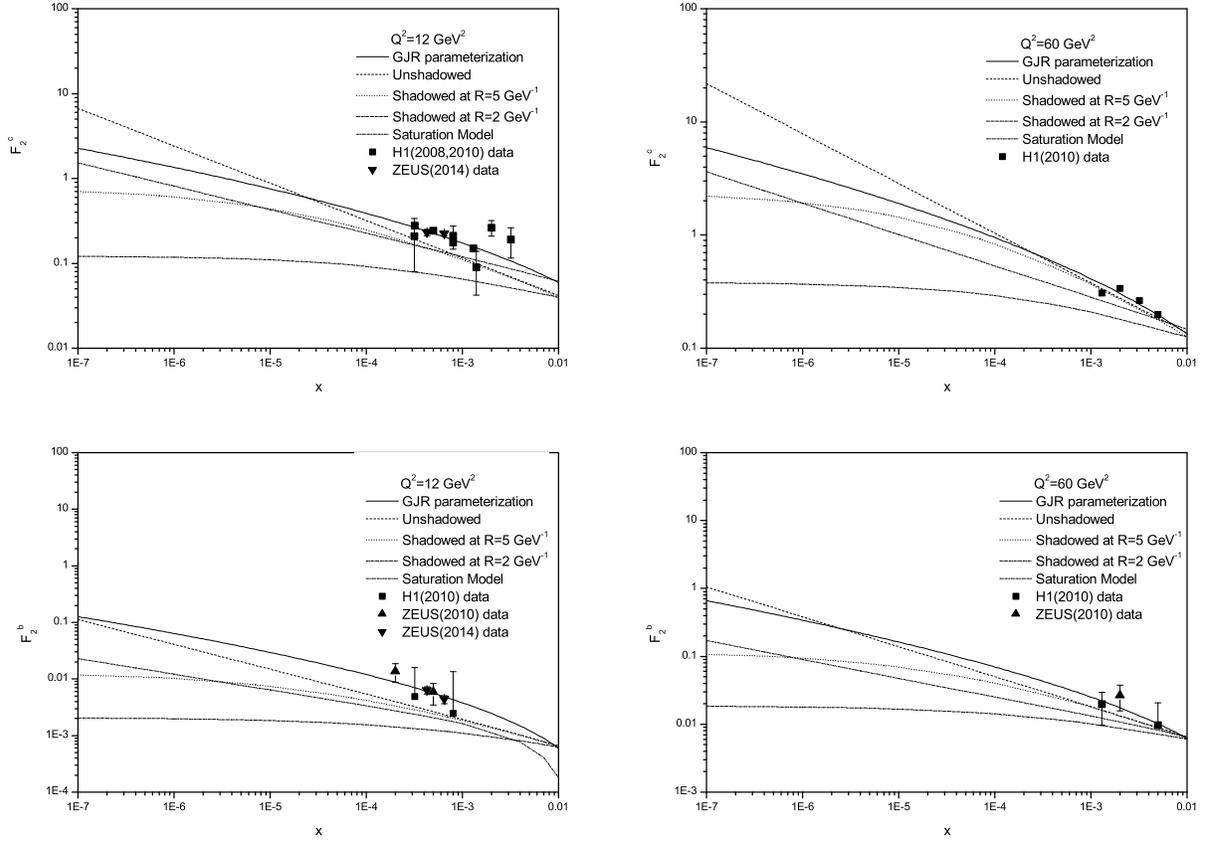}
\caption{ Heavy flavor structure functions $F_{2}^{i}(x,Q^{2})$ as
a function of the Bjorken variable $x$ at $Q^{2}$ values $12$ and
$60~GeV^{2}$ in comparison with the experimental data from ZEUS
and H1 Collaborations [11-15]  accompanied by total errors and GJR
parameterization [64]. The unshadowed and shadowed corrections and
also the saturation model concerning the geometrical scaling are
shown in this figure. }\label{Fig1}
\end{figure}
\begin{figure}
\includegraphics[width=1\textwidth]{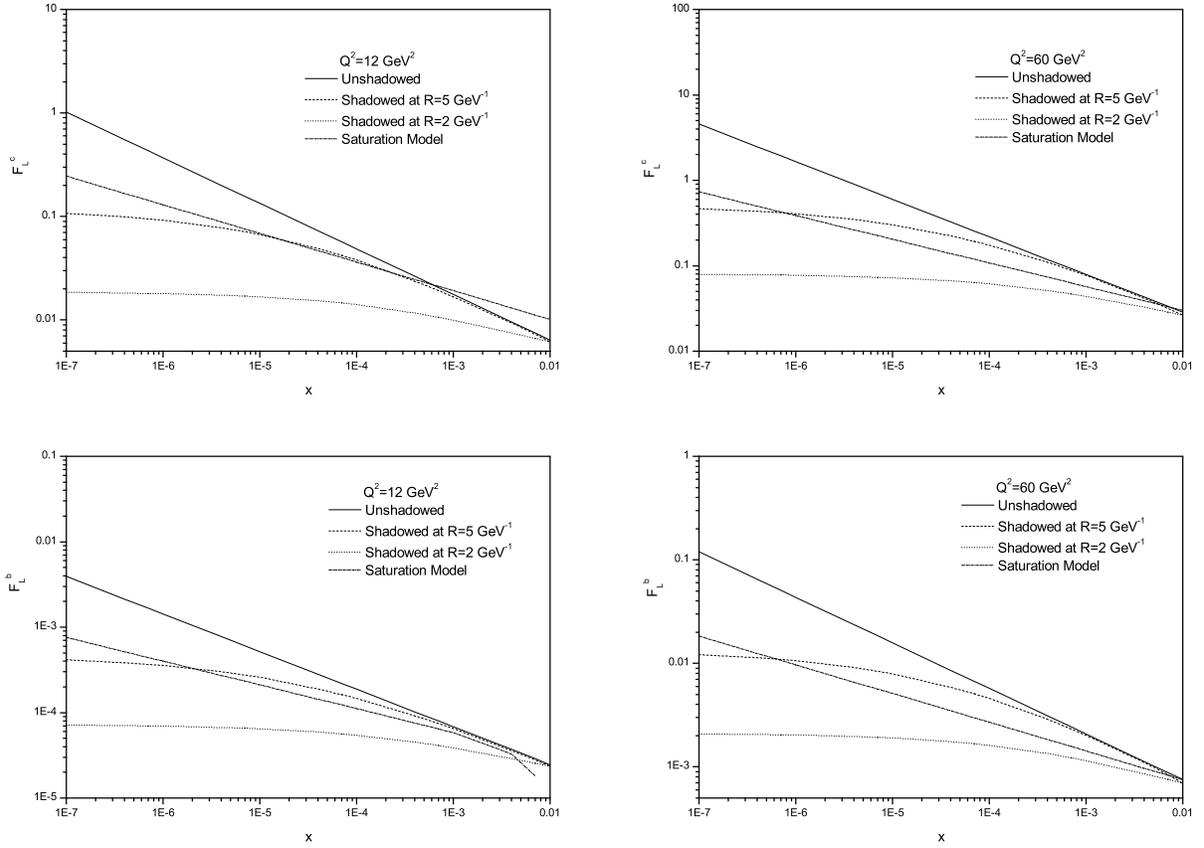}
\caption{Heavy flavor longitudinal structure functions
$F_{L}^{i}(x,Q^{2})$ as a function of the Bjorken variable $x$ at
$Q^{2}$ values $12$ and $60~GeV^{2}$. The unshadowed and shadowed
corrections and also the saturation model concerning the
geometrical scaling are shown in this figure.} \label{Fig2}
\end{figure}
\begin{figure}
\includegraphics[width=1\textwidth]{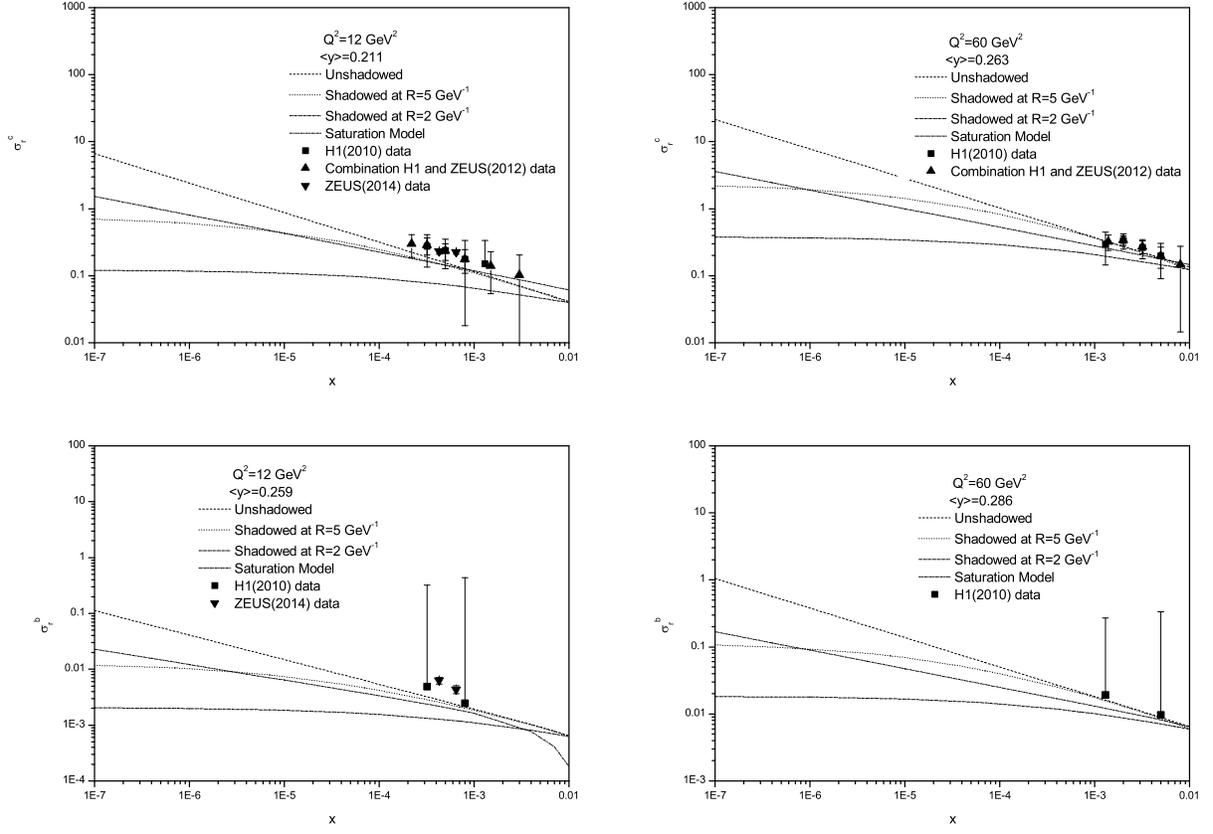}
\caption{Heavy flavor reduced cross section
$\sigma_{r}^{i}(x,Q^{2})$ as a function of the Bjorken variable
$x$ at $Q^{2}$ values $12$ and $60~GeV^{2}$ in comparison with the
experimental data from ZEUS and H1 Collaborations [11-15]
accompanied by total errors at inelasticity $<y>$. The unshadowed
and shadowed corrections and also the saturation model concerning
the geometrical scaling are shown in this figure.}\label{Fig3}
\end{figure}
\begin{figure}
\includegraphics[width=1\textwidth]{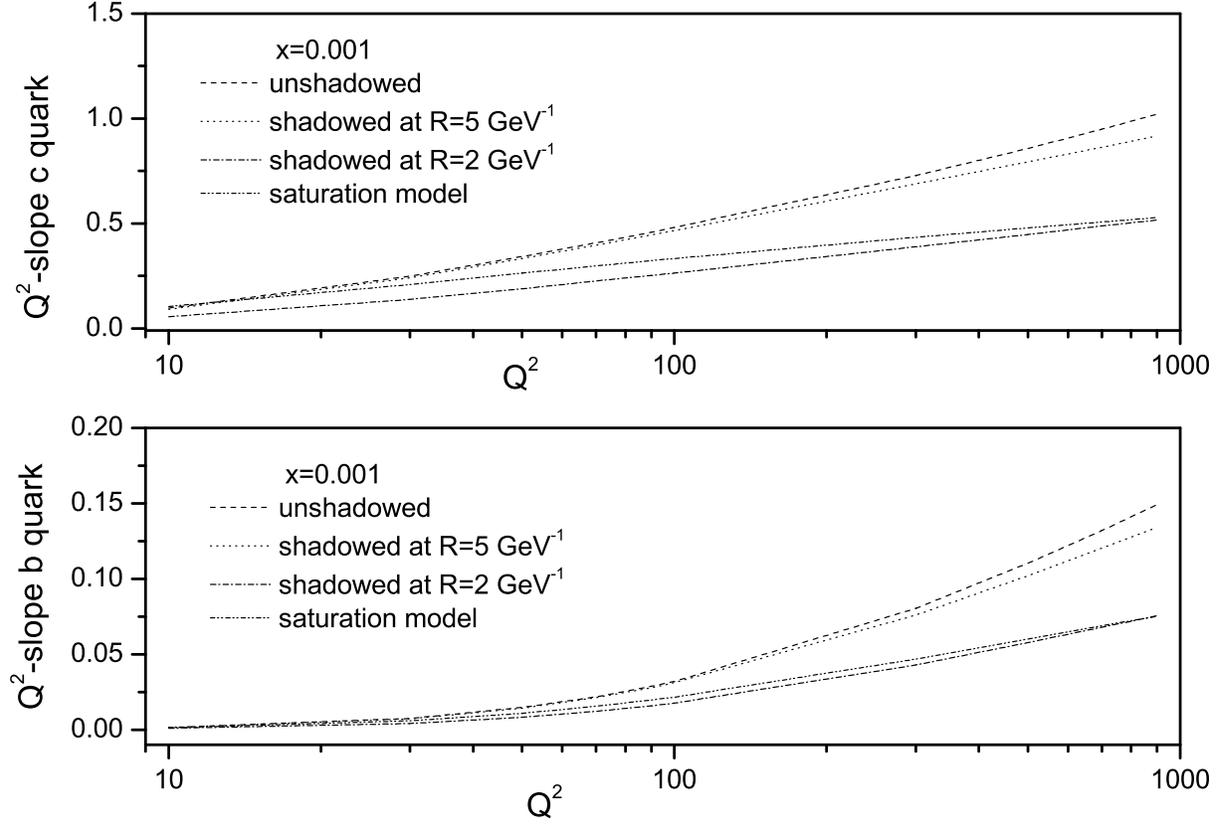}
\caption{$Q^{2}$-slope heavy flavor structure functions
($\partial{F_{2}^{i}}/\partial{\ln{Q^{2}}}$) as function of
$Q^{2}$ at $x=0.001$. Our results shown for unshadowed and
shadowed corrections and also for saturation model with respect to
the geometrical scaling in $Q^{2}$-slope heavy flavor structure
functions.}\label{Fig4}
\end{figure}
\begin{figure}
\includegraphics[width=1\textwidth]{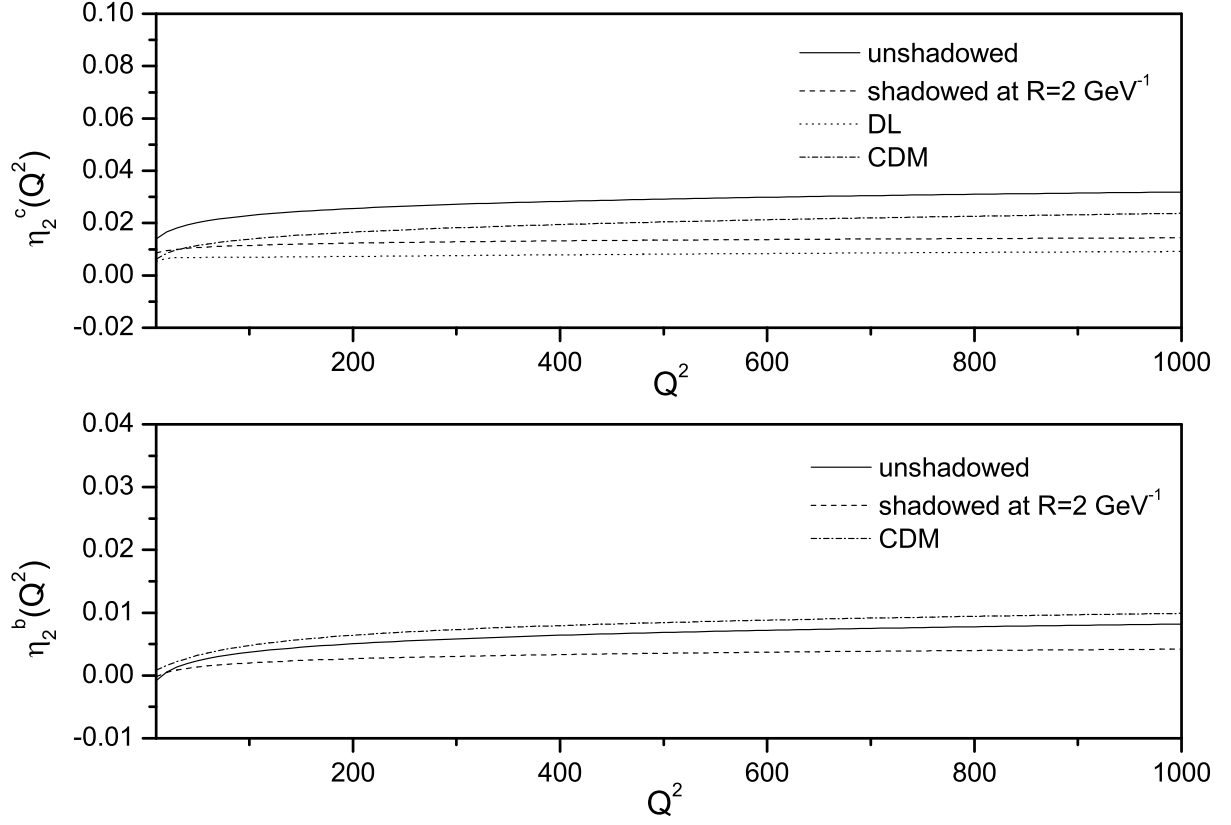}
\caption{The coefficients $\eta_{2}^{i}(Q^{2})$ from fits of the
form $F^{i}_{2}(x,Q^{2})=\eta_{2}^{i}(Q^{2})x^{-\lambda}$ to the
shadowed and unshadowed corrections compared to DL model [49-56]
and color dipole  model (CDM) [1-4].}\label{Fig5}
\end{figure}
\end{document}